\documentclass[12pt]{article}
\usepackage{amsfonts}
\usepackage{graphicx}
\usepackage{epstopdf}
\usepackage{epsfig}
\usepackage{amssymb}
\usepackage{setspace}
\usepackage{caption}
\usepackage{amsfonts}
\usepackage{color}
\usepackage{amsmath}
\usepackage{float}
\usepackage{comment}
\usepackage[utf8]{inputenc}
\newcommand {\be}{\begin{equation}}
\newcommand {\ee}{\end{equation}}
 \newcommand {\bea}{\begin{array}}
 
 \newcommand {\eea}{\end{array}}

\textwidth=6.5in \hoffset=-.75in \voffset=-1in \numberwithin{equation}{section}
\numberwithin{figure}{section}

\begin{document}

\begin{titlepage}
\vspace{1cm} 
\begin{center}
{\Large \bf {Kerr-Sen-Taub-NUT spacetime and circular geodesics}}\\
\end{center}
\vspace{2cm}
\begin{center}
\renewcommand{\thefootnote}{\fnsymbol{footnote}}
Haryanto M. Siahaan{\footnote{haryanto.siahaan@unpar.ac.id}}\\
Center for Theoretical Physics, Department of Physics,\\
Parahyangan Catholic University,\\
Jalan Ciumbuleuit 94, Bandung 40141, Indonesia
\renewcommand{\thefootnote}{\arabic{footnote}}
\end{center}

\begin{abstract}
We present a novel solution obeying classical equation of motion in the low energy limit of heterotic string theory. The solution represents a rotating mass with electric charge and gravitomagnetic monopole moment. The corresponding conserved charges are discussed. We also study the test body circular equatorial motions in the spacetime.
\end{abstract}
\end{titlepage}\onecolumn 
\bigskip 

\section{Introduction}
\label{sec:intro}

String theory offers a consistent quantum description of gravity and unification of all forces in nature. However, the lacking for experimental verifications yield the final answer for quantum gravity explanation is still opened. While waiting for proofs or insights from experimental aspects, theoretical works related to string theory constantly appear, including ones which make predictions from the low energy limit effective field theory.

In particular of our interest is the low energy limit effective field description of heterotic string theory. In this framework, Ashoke Sen has constructed a solution describing rotating and charged black holes, known as Kerr-Sen black holes, analogous to the Kerr-Newman black holes of Einstein-Maxwell theory. Despite the similarities between Kerr-Sen and Kerr-Newman solutions, there are several features which distinguish between the two solutions. For example, entropy of an extremal Kerr-Sen black hole can be expressed solely on its angular momentum \cite{Sen:1992ua}, the lacking of $Q$-picture hidden conformal symmetry for the generic Kerr-Sen spacetime \cite{Ghezelbash:2012qn,Chen:2010zwa}, and its unique behavior of light deflection \cite{Gyulchev:2006zg}. 

Subsequently Johnson and Myers discovered a static dyonic Taub-NUT solution of the low energy heterotic string \cite{Johnson:1994ek}, starting with the Taub-NUT solution to the vacuum Einstein equation as the seed metric. In their works, the authors of \cite{Sen:1992ua,Johnson:1994ek} employed the Hassan-Sen \cite{Hassan:1991mq} transformation that can map any stationary and axial symmetric spacetime which solves the vacuum Einstein equation to a new solution in the low energy limit of heterotic string theory. Recently, the author of \cite{Siahaan:2018qcw} employ the same method to obtain the acccelerating spacetime in the low energy limit of heterotic string using the $C$-metric as the seed solution. Interestingly, aspects of black holes in the low energy limit of heterotic string theory have been studied quite comprehensively in literature \cite{Duztas:2018adf,Duztas:2018ebr,Vieira:2018hij,Gwak:2016gwj,Uniyal:2017yll,Liu:2018vea,Siahaan:2015ljs,Siahaan:2015xna,Bernard:2017rcw}.

Taub-NUT solution in vacuum Einstein theory can be considered as an extension of  Schwarzschild solution, where in addition to the mass there exist a continuous NUT parameter $l$ which nowadays quite common to be referred as the magnetic mass or the gravitomagnetic monopole moment. The Taub-NUT solution can also be generalized to have an electric charge and rotation, known as the Kerr-Newman-Taub-NUT (KNTN) solution that solves the Einstein-Maxwell equations \cite{Griffiths:2009dfa}. The KNTN spacetime possesses the timelike and axial Killing symmetries, just like the Kerr or Kerr-Newman solutions. Recently, there appear a proposal where the gravitomagnetic monopole could be non-zero \cite{Chakraborty:2017nfu}, hence the Kerr-Taub-NUT (KTN) or KTNT solutions may have some physical relevances.

Since KTN spacetime solves the vacuum Einstein equations and has the timelike and axial symmetries, one may wonder the outcome of using this solution as a seed in the Hassan-Sen transformation. We could expect the result is something similar to the KNTN in Einstein-Maxwell theory, which is true and then can be called the Kerr-Sen-Taub-NUT (KSTN) solution. Some features of KNTN spacetime have been studied quite extensively in literature\cite{Chakraborty:2017nfu,Chakraborty:2019rna,Pradhan:2014zia,Sakti:2017pmt,Sakti:2019krw,Sakti:2017pmt,Aliev:2008wv,Duztas:2017lxk,Cebeci:2015fie,Zakria:2015eua,Mukherjee:2018dmm,Abdujabbarov:2011uv,Abdujabbarov:2012jj,Ahmedov:2011wb,Abdujabbarov:2008mz,Abdujabbarov:2012bn}, then investigating some features of KSTN spacetime can be potentially appealing. Some of our particular interests are the conserved charges associated to the solution and circular geodesics in KSTN spacetime.

The organization of this papers is as the followings. In section \ref{sec.Twisting}, we provide a quick review on the low energy heterotic string theory action and obtain the KSTN solution. In section \ref{sec.masscharge}, we obtain the electric charge, mass, and angular momentum in KSTN spacetime. Then, in section \ref{sec.CirGEO} we study the timelike and null circular geodesics in KSTN spacetime, followed by the accompanying effective potential plots in the next section. Finally, we present conclusions and discussions.

\section{Twisting the Kerr-Taub-NUT spacetime}\label{sec.Twisting}

\subsection{Symmetry of fields in low energy effective action}

In this paper we consider the stationary and axially symmetric spacetimes, $ g_{\mu \nu }  =  g_{\mu \nu } \left( {r,x} \right)$, with the coordinates\footnote{The coordinate $x=\cos\theta$ in Boyer-Lindquist type.} $x^\mu   = \left[ {t,r,x,\phi } \right]$. A set of fields $\left\{ {{G_{\mu \nu }},\Phi ,{A_\mu },{B_{\mu \nu }}} \right\}$ which solves the equations of motion derived from the four dimensional low energy effective action for heterotic string 
\be\label{action.het.string} S = \int {{d^4}x\sqrt { - G} {e^{ - \Phi }}\left( {R \left(G\right) + {{\nabla _\mu  \Phi \nabla ^\mu  \Phi }} - \frac{1}{8}{F_{\mu \nu }}{F^{\mu \nu }} - \frac{1}{{12}}{H_{\mu \nu \alpha }}{H^{\mu \nu \alpha }}} \right)} \ee
can been expressed in terms of the seed solution \cite{Siahaan:2018qcw}
\be \label{metric.seed. general}
ds^2  = \tilde g_{tt} dt^2  + 2\tilde g_{t\phi } dtd\phi  + \tilde g_{\phi \phi } d\phi ^2  + \tilde g_{rr} dr^2  + \tilde g_{xx} dx^2 \,.
\ee  
The seed metric $\tilde{g}_{\mu\nu}$ is stationary and axially symmetric solution to the vacuum Einstein equation. Indeed, the vacuum Einstein system is a special case of action (\ref{action.het.string}), i.e. when all fields but the spacetime metric vanish. Therefore, the Hassan-Sen transformation which maps the vacuum Einstein metric $\left\{ {\tilde g_{\mu \nu } } \right\}$ to a set of fields $\left\{ {{G_{\mu \nu }},\Phi ,{A_\mu },{B_{\mu \nu }}} \right\}$ can be simply considered as a mapping between solutions in the theory described by action (\ref{action.het.string}).  

Explicitly, the fields solutions $\left\{ {{G_{\mu \nu }},\Phi ,{A_\mu },{B_{\mu \nu }}} \right\}$ can be expressed as the followings. In the string frame, the spacetime metric reads
\be \label{metric.general.stringframe.sols}
{\rm{d}}s^2  = G_{\mu \nu } {\rm{d}}x^\mu  {\rm{d}}x^\nu   = \frac{{\tilde g_{tt} }}{{\Lambda ^2 }}\left( {{\rm{d}}t + \frac{{\tilde g_{t\phi } }}{{\tilde g_{tt} }} {\cosh^2 \left(\frac{\alpha}{2}\right)} {\rm{d}}\phi } \right)^2  + \tilde g_{rr} {\rm{d}}r^2  + \tilde g_{xx} {\rm{d}}x^2  + \left( {\tilde g_{\phi \phi }  - \frac{{\tilde g_{t\phi }^2 }}{{\tilde g_{tt} }}} \right){\rm{d}}\phi ^2 \,.
\ee 
where $\Lambda  = 1 + \left( {1 + {{\tilde g}_{tt}}} \right) {\sinh^2 \left(\alpha/2 \right)}$. The components of $U(1)$ gauge field are
\be\label{vectorU1.sols}
{A_t} = \frac{\sinh \left(\alpha \right)}{\Lambda }\left( {1 + {{\tilde g}_{tt}}} \right)~~{\rm and}~~{A_\varphi } = \frac{\sinh \left(\alpha \right)}{\Lambda }{{\tilde g}_{t\varphi }}\,,\ee
which give the field strength tensor ${F_{\mu \nu }} = {\partial _\mu }{A_\nu } - {\partial _\nu }{A_\mu }$ in action (\ref{action.het.string}) above. The dilaton field is 
\be\label{Dilaton.sols}
\Phi = -\ln\Lambda\,,
\ee 
and the non vanishing components of the second-rank antisymmetric tensor field are 
\be \label{Bmn.sols}
{B_{t\varphi }} =  - {B_{\varphi t}} = {\Lambda ^{ - 1}}{{\sinh^2 \left(\alpha/2 \right)}}{{\tilde g}_{t\varphi }}\,.
\ee 
The third-rank tensor appearing in action (\ref{action.het.string}) consists of the gauge field $A_\mu$ and tensor field $B_{\mu\nu}$, i.e.
\be {H_{\alpha \beta \mu }} = {\partial _\alpha }{B_{\beta \mu }} + {\partial _\mu }{B_{\alpha \beta }} + {\partial _\beta }{B_{\mu \alpha }} - \frac{1}{4}\left( {{A_\alpha }{F_{\beta \mu }} + {A_\mu }{F_{\alpha \beta }} + {A_\beta }{F_{\mu \alpha }}} \right)\,,\ee 
where${H_{\alpha \beta \mu }}$ is just a Chern-Simons term in the absence of $B_{\mu\nu}$.

All non-gravitational fields $\left\{ {\Phi ,A_\mu  ,B_{\mu \nu } } \right\}$ disappear as $\sinh \left(\alpha/2 \right) \to 0$, and the metric (\ref{metric.general.stringframe.sols}) reduces to the original seed solution (\ref{metric.seed. general}). The resulting metric (\ref{metric.general.stringframe.sols}) has the timelike and axial Killing vectors, $\partial_t$ and $\partial_{\varphi}$ respectively, as in the case of seed metric. The conserved quantities related to these two Killing symmetries are given in section \ref{sec.masscharge}. In the following sections, the geodesic calculations will be performed in Einstein frame where the corresponding action reads \cite{Johnson:1994ek}
\be \label{action.Einsteinframe}
S = \int {d^4 x\sqrt { - g} \left( {R\left( g \right) - \frac{1}{2}\nabla _\mu  \Phi \nabla ^\mu  \Phi  - \frac{{e^{ - \Phi } }}{8}F_{\mu \nu } F^{\mu \nu }  - \frac{{e^{ - 2\Phi } }}{{12}}H_{\mu \nu \lambda } H^{\mu \nu \lambda } } \right)} \,.
\ee 
The relation between string and Einstein frames metrics is given by $G_{\mu \nu }  = e^\Phi  g_{\mu \nu } $. 

\subsection{Kerr-Sen-Taub-NUT solution}\label{sec.KSTN}

In constructing a set of fields representing a massive rotating charged object equipped with NUT parameter in the low energy limit of heterotic string theory, namely the Kerr-Sen-Taub-NUT (KSTN) solution, we take the Kerr-NUT metric as the seed metric ${\tilde g}_{\mu\nu}$, 
\[
{\rm d}s_{KTN}^2  =  - \frac{{\Delta _r }}{{\rho ^2 }}\left( {{\rm{d}}t - \left( {a\Delta _x  + 2l\left( {1 - x} \right)} \right){\rm{d}}\phi } \right)^2 
\]
\be \label{metricKTN2}
+ \rho ^2 \left( {\frac{{{\rm{d}}r^2 }}{{\Delta _r }} + \frac{{{\rm{d}}x^2 }}{{\Delta _x }}} \right) + \frac{{\Delta _x }}{{\rho ^2 }}\left( {a{\rm{d}}t - \left( {r^2  + \left( {a + l} \right)^2 } \right){\rm{d}}\phi } \right)^2 \,,
\ee 
where $\Delta _r  = r^2  - 2mr + a^2  - l^2 $, $\Delta _x  = 1 - x^2 $, and $\rho ^2  = r^2  + \left( {l + ax} \right)^2 $. Taking $l\to 0$ in the last equation yields the Kerr metric. Spacetime (\ref{metricKTN2}) is stationary and axial symmetric, and it solves the vacuum Einstein equation. Using the prescription (\ref{metric.general.stringframe.sols}), we have
\[
{\rm{d}}s^2_{\rm string}  = \frac{{a^2 \Delta _x  - \Delta _r }}{{\rho ^2 \Lambda ^2 }}\left( {{\rm{d}}t + 2 \cosh^2 \left(\frac{\alpha}{2}\right)\frac{{\Delta _r l \left( 1-x \right) - a\Delta _x \left(mr + l^2 + al \right)}}{{a^2 \Delta _x  - \Delta _r }}{\rm{d}}\phi } \right)^2 
\]
\be\label{metric.KSNUT.string}
+ \rho ^2 \left( {\frac{{{\rm{d}}r^2 }}{{\Delta _r }} + \frac{{{\rm{d}}x^2 }}{{\Delta _x }}} +\frac{{ \Delta _x \Delta _r }}{{\Delta _r  - a^2 \Delta _x }}{\rm{d}}\phi ^2 \right)  \,,
\ee 
where
\be\label{Lambda}
\Lambda  = 1 + \frac{{2~ {{\sinh^2 \left(\alpha/2 \right)}} \left( {mr + l^2  + lax} \right)}}{{\rho ^2 }}\,.
\ee 
According to (\ref{Dilaton.sols}) and (\ref{vectorU1.sols}), the corresponding dilaton and vector fields are
\be 
\Phi  =  - \ln \left( {\frac{{\rho^2 +2~{{\sinh^2 \left(\alpha/2 \right)}} \left( {mr + l^2  + lax} \right)}}{{\rho ^2 }}} \right)\,,
\ee 
and
\be 
A_\mu  {\rm{d}}x^\mu   = \frac{{2 \sinh\left(\alpha\right) \left[ {\left( {mr + l^2  + lax} \right){\rm{d}}t + \left(l\Delta _r \left( {1 - x} \right) - a\Delta _x \left( {mr + l^2  + al} \right)\right){\rm{d}}\phi } \right]}}{{\rho ^2  + 2~{{\sinh^2 \left(\alpha/2 \right)}} \left( {mr + l^2  + lax} \right)}}\,,
\ee
respectively. Finally, the non-zero components of $B_{\mu\nu}$ are given by (\ref{Bmn.sols}), i.e.
\be
B_{t\phi }  =  - B_{\phi t}  = 2 \sinh^2 \left(\alpha/2 \right)\frac{{l\Delta _r \left( {1 - x} \right) - a\Delta _x \left( {mr + l^2  + al} \right)}}{{\rho ^2  + 2~{{\sinh^2 \left(\alpha/2 \right)}} \left( {mr + l^2  + lax} \right)}}\,.
\ee 
In the next section, we obtain the conserved charges with respect to $\partial_t$ and $\partial_\phi$ symmetries, i.e. mass and angular momentum measured at infinity, respectively. For KSTN spacetime, the mass is found to be 
\be
M = m\left( {1 + \sinh ^2 \left( {\frac{\alpha }{2}} \right)} \right)\,,
\ee 
just like in the case of Kerr-Sen black holes \cite{Sen:1992ua}. The charge is also the same as that of Kerr-Sen, 

The Einstein frame version of the metric (\ref{metric.KSNUT.string}) is the one that we use in the following section, which can be written as
\[
ds^2  = g_{\mu\nu} {\rm{d}}x^\mu  {\rm{d}}x^\nu  = -\frac{\Xi}{\varrho^2}\left( {{\rm{d}}t - \frac{{2M\left( {\tilde \Delta _r l\left( {1 - x} \right) - a\Delta _x \left( {\left( {M - b} \right)r + l^2  + al} \right)} \right)}}{{\left( {M - b} \right)\Xi }}{\rm{d}}\phi  } \right)^2 
\]
\be 
+ \varrho^2 \left( {\frac{{{\rm{d}}r^2 }}{{\tilde \Delta _r }} + \frac{{{\rm{d}}x^2 }}{{\Delta _x }} + \frac{{\tilde \Delta _r \Delta _x {\rm{d}}\phi ^2 }}{{\Xi }}} \right)
\ee \label{KerrSenNUTmetric}
where 
\be 
\varrho^2  = r\left( {r + 2b} \right) + \left( {l + ax} \right)^2  + \frac{{2bl\left( {l + ax} \right)}}{{M - b}} \,,
\ee 
\be 
\tilde \Delta _r  = r^2  - 2\left( {M - b} \right)r + a^2  - l^2\,,
\ee 
and
\be 
\Xi = r^2  - 2\left( {M - b} \right)r + a^2 x^2  - l^2\,.
\ee 
Note that we have expressed the equations above in terms of the mass $M$ and charge $Q$ of Kerr-Sen black holes\footnote{In section \ref{sec.masscharge} we obtain the conserved charges, including mass and electric charge, in Kerr-Sen-Taub-NUT geometry.}. Alternatively, the non-gravitational fields can be expressed as
\be 
A_\mu  dx^\mu   = \frac{{\sqrt 2 Q \left[ {\left( {alx + l^2  + \left( {M - b} \right)r} \right){\rm{d}}t + \left( {l\tilde \Delta _r \left( {1 - x} \right) - a\Delta _x \left( {\left( {M - b} \right)r + l^2  + al} \right)} \right){\rm{d}}\phi } \right]}}{{\varrho \left( {M - b} \right)}}\,,
\ee 
\be 
\Phi  =  - 2\ln \left( {\frac{{\varrho }}{\rho }} \right)\,,
\ee \label{DilatonSOL}
and
\be
B_{t\phi }  =  - B_{\phi t}  = \frac{{Q^2 \left( {l\tilde \Delta _r \left( {1 - x} \right) - a\Delta _x \left[ {\left( {M - b} \right)r + l^2  + al} \right]} \right)}}{{M\left( {M - b} \right)\varrho }}\,.
\ee  \label{2ndrankTensorSOL}
The set of fields solutions (\ref{KerrSenNUTmetric}) - (\ref{2ndrankTensorSOL}) reduce to that of Kerr-Sen \cite{Sen:1992ua} after setting $l=0$, and becomes the standard Kerr-NUT (\ref{metricKTN2}) in the absence of electric charge $Q$. 

It has been well known that Taub-NUT or Kerr-Taub-NUT solutions suffer conical singularities, which according Misner can be removed by introducing a periodic timelike coordinate in the stationary spacetime\footnote{Clearly the conical singularities or periodic timelike coordinate are not the ones that we may expect to experience in reality.} \cite{Griffiths:2009dfa}.
The seed metric (\ref{metricKTN2}) has the semi-infinite line singularity, where it is regular on the half axis $x=1$, i.e.
\[
\mathop {\lim }\limits_{x \to 1} \frac{{2\pi }}{{1 - x^2 }}\sqrt {\frac{{g_{\phi \phi } }}{{g_{xx} }}}  = 2\pi \,,
\]
but conical defect occurs at $x=-1$. Therefore, it is not surprising to find that the metric (\ref{KerrSenNUTmetric}) also suffers the same problem. Consequently, the conical defect yields a black hole is not a quite well defined object in the KSTN spacetime. Nevertheless, there exist quite a number of works in literature which keep using the black hole terminology in a spacetime with conical defect coming from the presence of NUT parameter \cite{Zakria:2015eua,Kerner:2006vu}. 

Regardless that we have a well defined description of black holes in KSTN spacetime or not, the celestial kinematics of objects in this geometry are still fascinating to be investigated since they might have some relevances to astronomical observations. In order to do this, we need to explore more the properties of this spacetime. Despite we cannot have an event horizon in normal sense living in this spacetime, one can still locate the radius which yields the metric (\ref{KerrSenNUTmetric}) to be singular, that is given by ${\tilde \Delta_r} = 0$. They are located at
\be \label{rpm}
r_ \pm   = M - b \pm \sqrt {\left( {M - b} \right)^2  + l^2  - a^2 } \,.
\ee 
If we allow a black hole can exist in this Kerr-Sen-Taub-NUT spacetime\footnote{This situation is similar to that in Kerr-Newman-Taub-NUT spacetime, where there is a radius which yields the metric to be singular, and one could have a horizon which give exactly Kerr-Newman black hole's horizon in the absence of NUT parameter.}, then the radius of that black hole horizon would be $r_+$. From eq. (\ref{rpm}), we can see the maximal rotation or extremality is achieved at $ a^2= \left( {M - b} \right)^2  + l^2  $, i.e. to maintain $r_\pm$ to be real valued. Violation of this bound leads to a production of the naked ring-singularity. Furthermore, for the KSTN spacetime, there also exist the static limit surface 
\be 
r_{{\rm{ergo}}}  = M - b \pm \sqrt {\left( {M - b} \right)^2  + l^2  - a^2 x^2 } \,,
\ee 
which is the outer radial solution to $g_{tt}=0$ and widely known as the outer region of ergosphere. 

The KSTN spacetime (\ref{KerrSenNUTmetric}) is stationary and axial symmetric, symmetries inherited from the seed solution. Therefore, the corresponding Killing vectors are $\zeta_{\left(t\right)} \equiv \partial_t$ and $\zeta_{\left(\phi\right)} \equiv \partial_\phi$, which hint the existence of two associated conserved quantities, namely the energy $E$ and angular momentum $L$. As it is expected in a rotating spacetime, one can also show the dragging effect in KSTN spacetime. The angular velocity of a stationary observer with constant $r$ and $x$ is given by
\be 
\Omega  =  - \frac{{g_{t\phi } }}{{g_{\phi \phi } }} = \frac{{\cosh ^4 \left( {\alpha /2} \right)\tilde g_{tt} \tilde g_{t\phi } }}{{\left( {\tilde g_{\phi \phi } \tilde g_{tt} \Lambda ^2  - \tilde g_{t\phi }^2 \left( {\Lambda ^2  - \cosh ^4 \left( {\alpha /2} \right)} \right)} \right)}}\,,
\ee
which at the horizon reduces to
\be
\Omega _ +   = \frac{{a\left( {M - b} \right)}}{{2M\left( {r_ +  \left( {M - b} \right) + l^2 } \right)}}\,.
\ee 
In the vanishing of the NUT parameter $l$ in the last equation, we have the angular velocity of the Kerr-Sen black hole horizon \cite{Sen:1992ua}
\be 
\Omega _{{\rm{KS,}} + }  = \frac{a}{{2M\left( {M - b + \sqrt {\left( {M - b} \right)^2  - a^2 } } \right)}}\,.
\ee 
The area covered by the sphere with radius $r_+$ can be computed as
\be \label{area}
{\cal A} = \int\limits_{0}^{2\pi} {\int\limits_{-1}^{1 } {\sqrt {g_{xx} g_{\phi \phi } } {\rm{d}}x{\rm{d}}\phi } }  = \frac{{8\pi M\left( {r_ +  \left( {M - b} \right) + l^2 } \right)}}{{M - b}}\,,
\ee 
which at $l\to 0$ is just the area of Kerr-Sen black hole.

Just like the analogous solution in Einstein-Maxwell theory, i.e. KNTN spacetime, the squared of Riemann tensor or Kretschmann scalar for KSTN spacetime is not singular at $r\to 0$. This property is due to the presence of NUT parameter $l$. In a spacetime where black holes can live, the singular Kretschmann scalar at $r\to 0$ is interpreted as the physical singularity that exist in the center of a black hole. Although a mathematical symbolic manipulation programs like MAPLE or Mathematica can do the computation to get the Kretschmann scalar for a spacetime, its full expression can be lengthy. This is exactly the case for KSTN spacetime, which is the geometry under investigation in this paper. Nevertheless, to support our future claim related to the curve of effective potential, let us just show the Kretschmann scalar for KSTN spacetime evaluated at $r=0$ and on equatorial plane. The reading is
\be\label{RR}
\left. {R^{\kappa \lambda \mu \nu } R_{\kappa \lambda \mu \nu } } \right|_{r = 0,x = 0}  = \frac{{4m^2 }}{{\left( {m + 2b} \right)^6 l^{12} }}\sum\limits_{k = 0}^4 {c_k l^{2k} } \,,
\ee
where 
\begin{flalign*}
&c_0 =4a^4 m^4 b^4 \,,\\
&c_1 ={8a^4 m^2 b^4  + 12m^4 b^4 a^2  + 16a^2 m^5 b^3 }\,,\\
&c_2 = {20m^6 b^2  + 36m^5 b^3  + b^2 \left( {16a^2  + 19b^2 } \right)m^4  + 48m^3 b^3 a^2  + 24m^2 b^4 a^2  + 4a^4 b^4 }\,,\\
&c_3 = {68m^4 b^2  + 88m^3 b^3  + \left( {16b^2 a^2  + 38b^4 } \right)m^2  + 32a^2 b^3 m + 12a^2 b^4  - 12m^6 }\,,\\
&c_4 = {12m^4  + 19b^4  + 48m^3 b + 72m^2 b^2  + 52b^3 m}\,.
\end{flalign*}
In expressions above, $m=M-b$, and we prefer to use $m$ instead of $M-b$ for economical reason. Note that the squared of Riemann tensor $R^{\kappa \lambda \mu \nu } R_{\kappa \lambda \mu \nu }$ evaluated at $r=0$ computed in (\ref{RR}) is finite for a non-vanishing $l$.

\section{Mass and angular momentum}\label{sec.masscharge}

Defining mass and angular momentum in a spacetime with NUT parameter is not quite simple. In an asymptotically flat spacetime, for example Kerr-Newman family without NUT parameter in Einstein-Maxwell theory, one can employ the Komar integral to get the mass and angular momentum. However, when the NUT parameter is present, one needs to carefully treat the surface integral at infinity to get the desired conserved charges associate to the $\partial_t$ and $\partial_\phi$ Killing vectors\cite{Aliev:2008wv}. 

Under some general diffeomorphisms, we can employ the Barnich-Brandt method \cite{Barnich:2001jy} to compute the conserved charges in any field theory, including those containing gravity. In Kerr/CFT correspondence, this method is used to obtain charges in the theory under consideration \cite{Guica:2008mu,Compere:2012jk,Ghezelbash:2009gf}. Virasoro algebra between these charges contains the central charge which is crucial in reproducing the Bekenstein-Hawking entropy for black holes in the spacetime whose asymptotic symmetries are associated to the obtained central charge. Clearly, the Barnich-Brandt method can also be applied for the exact diffeomorphisms $\partial_t$ and $\partial_\phi$.

The low energy limit of heterotic string discussed in this paper contains four fields, where there exist diffeomorphisms associated to each of these fields. Consequently, each of these diffeomorphisms may contribute to the central charge according to Barnich-Brandt method. In this section, we compute the conserved charges associated to $\partial_t$ and $\partial_\phi$ Killing vectors of KSTN spacetime using the Barnich-Brandt method. As the starting point, let us reparameterize the mass $m$ as one parameter family $m = s {\bar m}$, where $s \in \left[ {0,1} \right]$. In such consideration, one can define the variation of each fields as follows\footnote{At the end of calculation, we can perform the integration $\int\limits_0^1 {ds} $ for the obtained conserved charges.}
\be 
h_{\mu \nu }  = \frac{{dg_{\mu \nu } }}{{ds}}ds~~,~~a_\mu   = \frac{{dA_\mu  }}{{ds}}ds~~,~~b_{\mu \nu}  = \frac{{dB_{\mu \nu} }}{{ds}}ds~~,~~\phi  = \frac{{d\Phi }}{{ds}}ds\,.
\ee 
Accordingly, we have 
\be 
f_{\mu \nu }  = \nabla _\mu  a_\nu   - \nabla _\nu  a_\mu \,,
\ee 
and
\be 
\tilde h_{\mu \nu \rho }  = \partial _\mu  b_{\nu \rho }  + \partial _\rho  b_{\mu \nu }  + \partial _\nu  b_{\rho \mu }  - \frac{1}{4}\left( {a_\mu  f_{\nu \rho }  + a_\rho  f_{\mu \nu }  + a_\nu  f_{\rho \mu } } \right)\,.
\ee 

The charge formula as a result of diffeomorphisms of each fields read
\be\label{chargeBB} 
Q_{\zeta}    =\frac{1}{{8\pi }}\int\limits_0^1 {ds} \oint_S {\left( {k_{\left( {\zeta ;g} \right)}  + k_{\left( {\zeta ;A} \right)}  + k_{\left( {\zeta ;\Phi } \right)}  + k_{\left( {\zeta ;B} \right)} } \right)}\,,
\ee 
where
\be \label{kg}
k_{\left( {\zeta ;g} \right)}^{\alpha \beta }= \left\{\zeta ^\beta  \nabla _\mu  h^{\alpha \mu }  - \zeta ^\beta  \nabla ^\alpha  h - \zeta _\mu  \nabla ^\beta  h^{\alpha \mu }  - \frac{h}{2}\nabla ^\beta  \zeta ^\alpha   + h^{\beta \mu } \nabla _\mu  \zeta ^\alpha   - \frac{{h^{\mu \beta } }}{2}\left( {\nabla ^\alpha  \zeta _\mu   + \nabla _\mu  \zeta ^\alpha  } \right)\right\}\,,
\ee 
\be \label{kA}
k_{\left( {\zeta ;A} \right)}^{\mu \nu }   = \left\{ {\left( {F^{\mu \rho } h_\rho ^\nu   - \frac{{f^{\mu \nu } }}{2} - \frac{{hF^{\mu \nu } }}{4}} \right)\zeta ^\alpha  A_\alpha   - \frac{{F^{\mu \nu } }}{2}\zeta ^\alpha  a_\alpha   - F^{\alpha \mu } \zeta ^\nu  a_\alpha   - \frac{{a^\mu  }}{2}\left( {\zeta _\alpha  F^{\nu \alpha }  + \partial ^\nu  \left( {\zeta ^\alpha  A_\alpha  } \right)} \right)} \right\} \,,
\ee 
\be \label{kPhi}
k_{\left( {\zeta ;\Phi } \right)}^{\mu \nu }  =  - \frac{2}{3}\phi \zeta ^\mu  \partial ^\nu  \Phi dS_{\mu \nu }\,,
\ee 
and
\[
k_{\left( {\zeta ;B} \right)}  = \frac{1}{{12}}\zeta ^\lambda  \left( {\varepsilon _{\mu \nu \rho \beta } b_{\lambda \alpha }  + \varepsilon _{\mu \nu \rho \alpha } b_{\beta \lambda }  + \varepsilon _{\mu \nu \rho \lambda } b_{\alpha \beta } } \right)H^{\mu \nu \rho } dx^\alpha   \wedge dx^\beta  
\]
\be\label{kB}
+ \left\{ {\frac{1}{3}\left( {\zeta ^\lambda  b_{\alpha \lambda } H^{\mu \alpha \rho }  + \zeta ^\lambda  B_{\alpha \lambda } \left( {\tilde h^{\mu \alpha \rho }  + \frac{h}{2}H^{\mu \alpha \rho } } \right)} \right) + \frac{1}{2}b^{\mu \alpha } \left( {B^{\nu \beta } \partial _\alpha  \zeta _\beta   + B_{\rho \alpha } \partial ^\nu  \zeta ^\rho   + \zeta ^\rho  \partial _\rho  B^\nu  _\alpha  } \right)} \right\}dS_{\mu \nu } \,.
\ee
In equation above, $k_{\left( {\zeta ;\Psi } \right)}  = k_{\left( {\zeta ;\Psi } \right)}^{\mu \nu } dS_{\mu \nu } $ where $\Psi$ is the fields $\left\{ {g,A,\Phi } \right\}$, and we have used 
\be\label{2sphere} 
dS_{\mu \nu }  = \frac{1}{4}\varepsilon _{\mu \nu \alpha \beta } dx^\alpha   \wedge dx^\beta  \,,
\ee 
as the two-surface orthonormal to the timelike and radial directions with $\left| {\varepsilon _{0123} } \right| = \sqrt { - g} $. In writing equations (\ref{kg}) - (\ref{kB}), we have omitted the contributions from the gauge freedoms of $A_\mu$ and $B_{\mu\nu}$ fields. The mass can be computed by taking $\zeta = \zeta_{\left(t\right)}$ in (\ref{chargeBB}), which gives
\be 
Q_{\zeta _{\left( t \right)} }   = m\left( {s^2+ 1} \right)\,.
\ee 
This is exactly the mass of Kerr-Sen black holes \cite{Sen:1992ua,Ghezelbash:2009gf}, which can be computed using the standard Komar integral. In fact, the same mass can also be obtained using Abbott-Deser-Tekin method \cite{Peng:2016wzr}. Furthermore, in calculation above, the conserved mass is given solely from the gravity contribution, i.e. $k_{\left( {\zeta ;g} \right)}$ integration. 

On the other hand, employing the charge formula for the axial Killing vector $\zeta _{\left( \phi  \right)}  $ yields
\be 
Q_{\zeta _{\left( \phi  \right)} }  = m\left( {1 + s^2 } \right)\left( {a + 3l\left( {1 + s^2 } \right)} \right)\,.
\ee 
As one would expect, the last formula matches the angular momentum of Kerr-Sen black hole if one sets $l\to 0$. In other words, the angular momentum measured at infinity in KSTN spacetime gets contribution from the NUT parameter, which resembles the similar situation in KNTN case \cite{Aliev:2007fy}, and in the boosted Kerr-Taub-NUT study \cite{Aliev:2008wv}. To compute the electric charge, one can use the formula
\be\label{electric.charge}
Q = \frac{1}{4\pi} \mathop {\lim }\limits_{r \to \infty } \oint_S {F^{\mu \nu } dS_{\mu \nu } } = \sqrt 2 m\sinh \left( {\frac{\alpha }{2}} \right)\cosh \left( {\frac{\alpha }{2}} \right)\,,
\ee
which gives the same electric charge to that of Kerr-Sen solution \cite{Sen:1992ua}. Note that the electric charge above is obtained in the Einstein frame described by the action (\ref{action.Einsteinframe}). 

\section{Equatorial circular geodesics}\label{sec.CirGEO}

\subsection{Generalities}

In this section, we present some investigations on the equatorial and circular geodesics in the Kerr-Sen-Taub-NUT spacetime. As pointed out in \cite{Jefremov:2016dpi}, equatorial geodesics studies are not too straightforward for the spacetime with NUT parameter. In addition to the requirement $\dot r = 0$ and $\ddot r = 0$ for the circular motion, we also need to verify that there is no velocity and acceleration in $x$ direction\footnote{Or in $\theta$ direction for the standard Boyer-Lindquist type coordinate.} . In \cite{Mukherjee:2018dmm}, the authors employ the Hamilton-Jacobi equation in Kerr-Newman-NUT spacetime to study $\dot \theta =0$ and  $\ddot \theta=0$. 

In this paper, we prefer to compute $\dot x$ and $\ddot x$ by using the effective potential approach from the test body geodesics, just as in the same fashion we deal with the quantities $\dot r$ and $\ddot r$.  As a start, we consider a general Lagrangian of a neutral test particle living in a curved background 
\be \label{Lagrangian}
2 {\cal L} = g_{tt} \dot t^2  + 2g_{t\phi } \dot t\dot \phi  + g_{\phi \phi } \dot \phi ^2  + g_{rr} \dot r^2 + g_{xx} {\dot x}^2 \,,
\ee 
where the spacetime metric is a function of $r$ and $x$. Constants of motion related to this Lagrangian are
\be 
- E = g_{tt} \dot t + g_{t\phi } \dot \phi~~{\rm and}~~L = g_{\phi \phi } \dot \phi + g_{t\phi } \dot t \,.
\ee 
In the last equations, $E$ and $L$ are interpreted as energy and angular momentum of a test particle, respectively. Furthermore, a little algebra on the last equations yields
\be \label{dotTdotPhi}
\dot t \Delta_x {\Delta_r}  = g_{t\phi } L + g_{\phi \phi } E ~~{\rm and}~~ - \dot \phi \Delta_x {\Delta_r}  = g_{tt} L + g_{t\phi } E\,.
\ee 
Note that, in the last equation we have used a general relation that applies to Kerr-Sen-Taub-NUT spacetime, i.e. $g_{tt} g_{\phi \phi }  - g_{t\phi }^2  = \Delta _x \Delta _r $. If one considers an equatorial motion, then this relation reduces to the useful identity that normally employed in equatorial circular geodesics, namely $g_{tt} g_{\phi \phi }  - g_{t\phi }^2  = \Delta _r $.

In the followings, we look for some setups which allow circular geodesics on equatorial plane. From the metric that associate to the Lagrangian (\ref{Lagrangian}), we can have
\be \label{geo.gen}
\Theta  + g_{xx} \dot x^2  + g_{rr} \dot r^2  = \delta ,
\ee 
where $\Theta  \equiv g_{rr} \dot t^2  + 2g_{t\phi } \dot t\dot \phi  + g_{\phi \phi } \dot \phi ^2 $. Alternatively, one can show
\be \label{ThetaINEandL}
\Theta  = \frac{{g_{\phi \phi } E^2  + 2g_{t\phi } LE + g_{tt} L^2 }}{{\Delta _x \Delta _r }}
\ee 
after making use of (\ref{dotTdotPhi}). The metric signature that we are using yields $\delta = 0$ for null case and $\delta = -1$ for the timelike. Now let us first consider test body motions on equatorial plane, i.e. $x=0$, hence eq. (\ref{geo.gen}) now reads\footnote{Here $\left. X \right|_y $ means a function $X$ evaluated at some fixed $y$.}
\be \label{eqtr.geo}
\left. \Theta  \right|_{x=0}   + \left. g_{rr}  \right|_{x=0}  \dot r^2  = \delta \,.
\ee 
Accordingly, we can introduce an effective potential $V_{r,{\rm{eff}}} = -{\dot r}^2$ associated to (\ref{eqtr.geo}), namely
\be \label{Vgenr}
 \left. V_{r,{\rm{eff}}}  \right|_{x=0}  = \Delta _r \left. \frac{{\Theta  - \delta }}{{{\varrho} ^2 }} \right|_{x=0}  \,.
\ee 
A circular motion denoted by $\dot r=0$ requires this potential to be vanished. One small subtlety that needs attention here is to make sure the non-vanishing of
\be \label{rho0}
\varrho ^2 \left( {x = 0} \right) = r\left( {r + 2b} \right) + l^2  + \frac{{2bl^2 }}{{M - b}} \equiv \varrho _0^2 \,,
\ee 
hence the effective potential (\ref{Vgenr}) cannot be singular. However, guided by the physical insight that it is very unlikely for a massive body to maintain a huge amount of electric charge, namely $b < M$, we can safely claim the non-vanishing of (\ref{rho0}).

Similarly, we can also define an effective potential $V_{x,{\rm{eff}}} = -{\dot x}^2$ that coming from a fixed radius consideration in (\ref{geo.gen}). Instead of having (\ref{eqtr.geo}) as a start, here we have 
\be \label{cir.geo}
\Theta  + g_{xx} \dot x^2  = \delta \,,
\ee  
that leads us to
\be \label{Vgenx}
\left. V_{x,{\rm{eff}}} \right|_{r=r_0}  = \Delta _x \left. \frac{{\Theta  - \delta }}{{\varrho ^2 }}  \right|_{r=r_0} \,.
\ee 
Provided that $ \left. \Delta _r \right|_{r=r_0}\ne 0$, which is guaranteed by our interest in studying circular geodesics with radii $r_0$ outside horizon, the vanishing $\left. V_{x,{\rm{eff}}} \right|_{r=r_0}$ on equatorial plane can be achieved for the particular $r = r_0 \left(E,L,M,b,l\right)$ that solves $\left. V_{r,{\rm{eff}}}  \right|_{x=0}  = 0$ in eq. (\ref{Vgenr}). Furthermore, to ensure that there is no acceleration in $r$ or $x$ direction, the following constraints 
\be\label{dVdr} 
\frac{{dV_{r,{\rm eff}} }}{{dr}} = 0\,,\ee 
and
\be \label{dVdx}
\frac{{dV_{x,{\rm eff}} }}{{dx}} = 0\,,
\ee 
must be fulfilled as well. The last equation is normally satisfied in most spacetimes without NUT parameter, for example those in Einstein-Maxwell-dilaton \cite{Jai-akson:2017ldo} and low energy limit of heterotic string \cite{Siahaan:2019oik} theories. However, when NUT parameter exists, eq. (\ref{dVdx}) could add another constrain to the familiar ones obtained from $V_r$. Particularly for timelike case, the stability of circular geodesics is dictated by the behavior of second derivative $V_{r,{\rm eff}}$ with respect to radius, and to find the so called innermost stable circular orbit (ISCO) we make use the condition
\be\label{d2Vdr2} 
\frac{{d^2V_{r,{\rm eff}} }}{{dr^2}} = 0\,.\ee 
In the followings, we employ the generalities that have been developed here to verify the existence of equatorial circular orbits in both timelike and null cases. 

\subsection{Circular timelike orbits}

In timelike consideration, from eq. (\ref{dVdx}) we can have
\be \label{dVdxTL}
\left\{ \left( {4Ml + 2\left( {M - b} \right)a} \right)E^2  - 2L\left( {M - b} \right)E + a\left( {M - b} \right) \right\} l = 0\,.
\ee 
This equation is satisfied for zero NUT parameter, or the term inside the curly bracket vanish. The first case is the reason why equatorial geodesics in Kerr-Sen spacetime is guaranteed \cite{Siahaan:2015ljs}. If we are interested in the non-vanishing of NUT parameter $L$, then the energy and angular momentum of test particles get an additional constraint from the term in curly bracket above. Using this condition to constrain the angular momentum $L$ yields the reading of $V_{r,{\rm{eff}}} = 0$, $V'_{r,{\rm{eff}}} = 0$, and $V''_{r,{\rm{eff}}} = 0$ equations can be expressed as
\be \label{eq.ISCO-V}
f_{1} l^4  + f_{2} l^2  + f_{3}  = 0\,,
\ee 
\be\label{eq.ISCO-dV}
f_{4} l^4  + f_{5} l^2  + f_{6}  = 0\,,
\ee
\be \label{eq.ISCO-ddV}
f_{7} l^4  + f_{8} l^2  + f_{9}  = 0\,,
\ee 
where
\[
f_1  = 4E^2 \left( {M + b} \right)\left\{ {M\left( {E^2  - 1} \right) + b\left( {E^2  + 1} \right)} \right\}\,,~~~~~~~~~~~~~~~~~~~~~~~~~~~~~~~~~~~~~~~~~~~~~~~~~~~~~~~~~~~~
\]
\[
f_2  =  - 8E^2 r\left( {M - b} \right)^3  + \left( {8E^2 r\left\{ {E^2 \left( {r + 2b} \right) - 3b} \right\} + a^2 } \right)\left( {M - b} \right)^2  + 8E^2 br\left( {E^2 \left\{ {2r + b} \right\} + r} \right)\left( {M - b} \right)\,,
\]
\[
f_3  = 2r\left( {a^2  - 4E^2 \left\{ {r^2  + 2b} \right\}} \right)\left( {M - b} \right)^3  - r^2 \left( {a^2  - 4E^2 \left\{ {r + 2b} \right\}\left\{ {E^2 \left( {r + 2b} \right) + r} \right\}} \right)\left( {M - b} \right)^2 \,,~~
\]
\[
f_4  = 2E^2 \left( {E^2  - 2} \right)\left( {M - b} \right)^3  + 2E^2 \left( {E^2 \left\{ {r + 6b} \right\} + r - 5b} \right)\left( {M - b} \right)^2  + 8E^2 b\left( {2E^2 b + r} \right)\left( {M - b} \right) - 8E^4 b^2 r\,,
\]
\[
f_5  =  - 8E^2 r\left( {M - b} \right)^4  + \left( {a^2  + 4E^2 r\left\{ {E^2 \left( {4b + 3r} \right) - 8b} \right\}} \right)\left( {M - b} \right)^3 ~~~~~~~~~~~~~~~~~~~~~~~~~~~~~~~~~
\]
\[
~~~~~~~~~~+ r\left( {4E^2 r\left\{ {E^2 \left( {10b^2  + 6rb - r^2 } \right) + r^2  + 9rb} \right\} - a^2 } \right)\left( {M - b} \right)^2  - 8E^2 br^2 \left( {r + E^2 \left\{ {2r + 3b} \right\}} \right)\left( {M - b} \right)\,,
\]
\[
f_6  = \left( {2ra^2  - 8E^2 r^2 \left\{ {3b + 2r} \right\}} \right)\left( {M - b} \right)^4  + r^2 \left( {2E^2 \left\{ {E^2 \left( {16rb + 12b^2  + 5r^2 } \right) + 16rb + 10r^2 } \right\} - 3a^2 } \right)\left( {M - b} \right)^3 
\]
\[
+ \left( {a^2 r^3  - 2r^3 E^2 \left\{ {E^2 \left( {8b^2  + 10rb + 3r^2 } \right) + 3r^2  + 5rb} \right\}} \right)\left( {M - b} \right)^2 \,,~~~~~~~~~~~~~~~~~~~~~~~
\]
\[
f_7  = 2E^2 \left( {E^2  + 1} \right)\left( {M - b} \right)^2  + 8E^2 b\left( {M - b - E^2 b} \right)\,,~~~~~~~~~~~~~~~~~~~~~~~~~~~~~~~~~~~~~~~~~~~~~~~~~~~~~~~~~
\]
\[
f_8  =  - 8E^2 \left( {M - b} \right)^4  + \left( {8E^4 \left\{ {3r + 2b} \right\} - 32E^2 b} \right)\left( {M - b} \right)^3 ~~~~~~~~~~~~~~~~~~~~~~~~~~~~~~~~~~~~~~~~~~~~~~~~
\]
\[
~~~~~~~~~~+ \left( {4E^2 \left\{ {E^2 \left( {12rb - 3r^2  + 10b^2 } \right) + 18rb + 3r^2 } \right\} - a^2 } \right)\left( {M - b} \right)^2  - 24E^2 rb\left( {2E^2 \left\{ {r + b} \right\} + r} \right)\left( {M - b} \right)\,,
\]
\[
f_9  = \left( {2a^2  - 48E^2 r\left\{ {r + b} \right\}} \right)\left( {M - b} \right)^4  + 2r\left( {E^4 \left\{ {24b\left( {2r + b} \right) + 20r^2 } \right\} + 8rE^2 \left\{ {5r + 6b} \right\} - 3a^2 } \right)\left( {M - b} \right)^3 
\]
\[
+ r^2 \left( {3a^2  - E^4 \left\{ {48b^2  + 80rb + 30r^2 } \right\} - 10rE^2 \left\{ {3r + 4b} \right\}} \right)\left( {M - b} \right)^2\,. ~~~~~~~~~~~~~~~~~~~
\]

\begin{figure}
	\begin{center}
		\includegraphics[scale=0.6]{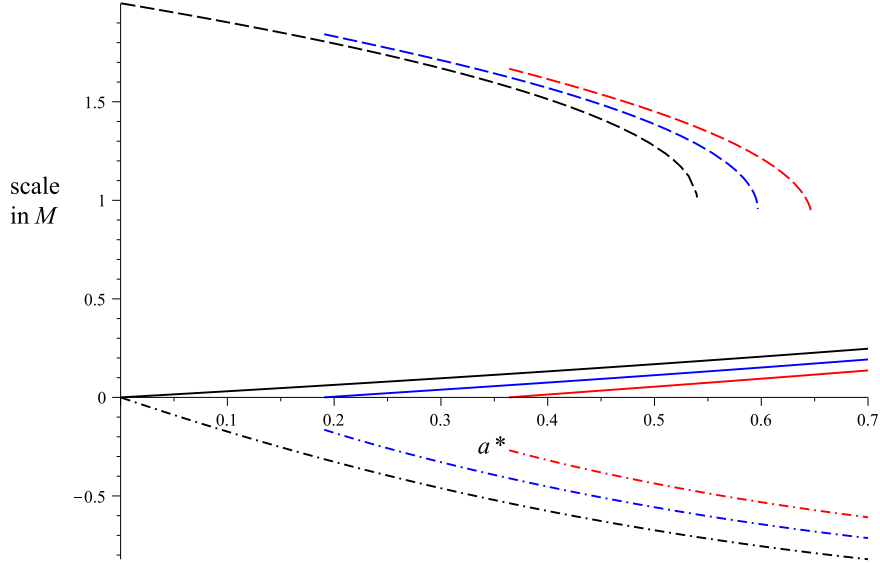}
	\end{center}
	\caption{Plots of ISCO radius (solid) in co-rotating case, outer horizons (dash), and the corresponding squared NUT parameter $l$ (dash-dot) that solve eqs. (\ref{eq.ISCO-V}), (\ref{eq.ISCO-dV}), and (\ref{eq.ISCO-ddV}). Plots in black color represent case of $b=0$, blue for $b=0.05~M$, and red for $b=0.1~M$. We have used $a^*$ to denote the ratio $a/M$.} \label{fig.ISCO}
\end{figure}

It is not easy to extract some qualitative results from these there equations, (\ref{eq.ISCO-V}), (\ref{eq.ISCO-dV}), and (\ref{eq.ISCO-ddV}). However, we can obtain some numerical plots from these three equations, which are presented in fig. \ref{fig.ISCO}. From these numerical results, we learn that not only the outcome for ISCO radius is smaller than the corresponding horizons, the associated NUT parameter $l$ is also purely imaginary, i.e. $l^2 < 0$. The case of timelike circular geodesics in Kerr-Taub-NUT spacetime is represented by the plots for $b=0$, which cannot take place on equatorial plane $x=0$. This is in agreement to a result in \cite{Jefremov:2016dpi}, which states that the circular motion in spacetime with NUT parameter is not in equatorial plane. Furthermore, from the numerical evaluations of $b=0.05~M$, and $b=0.1~M$, the same conclusions can also be drawn, namely the timelike circular geodesics cannot take place on the plane $x=0$ if the NUT parameter is non-zero. More numerical evaluations of for particular $b$ would likely to possess the same features, leading us to the same conclusion where circular timelike geodesics cannot exist on equatorial plane in Kerr-Sen-Taub-NUT spacetime.

\subsection{Light ring}

Now let us turn our discussion to the null case. Similarly, eq. (\ref{dVdx}) in null consideration 
\be 
\left\{ {\left( {\kappa  - a} \right)\left( {M - b} \right) - 2Ml} \right\} l =0\,,
\ee 
demands a constraint for $\kappa \equiv E^{-1} L$ if we let the NUT parameter to be non-zero. Solving an equation obtained from the term in curly bracket for $\kappa$ and substituting the result to $V_{\rm eff}=0$ and $V'_{\rm eff}=0$ give us
\be 
\Xi_0 \varrho_0^4 =0\,,
\ee 
and
\be
\Xi _0 \left\{ {\left( {M\left\{ {M + r + 2b} \right\} - 3b\left\{ {r + b} \right\}} \right)l^2  + r\left( {M - b} \right)\left( {M\left\{ {5r + 6b} \right\} - 3\left\{ {r + b} \right\}\left\{ {r + 2b} \right\}} \right)} \right\} = 0\,,
\ee 
respectively. In equations above, $\varrho_0\equiv\varrho\left(x=0\right)$ and $\Xi_0\equiv\Xi\left(x=0\right)$ appearing in the metric (\ref{KerrSenNUTmetric}). Here we consider non-zero $\varrho_0$ to maintain the effective potentials $V_{x,{\rm eff}}$ and $V_{r,{\rm eff}}$ to be non-singular. Nevertheless, one cannot get a real valued non-zero solution for the NUT parameter $l$ from the last two equations. Therefore we can conclude that light ring does not exist on equatorial plane in Kerr-Sen-Taub-NUT spacetime. 

\section{Effective potentials}\label{sec.effV}

\begin{figure}
	\begin{center}
		\includegraphics[scale=0.4]{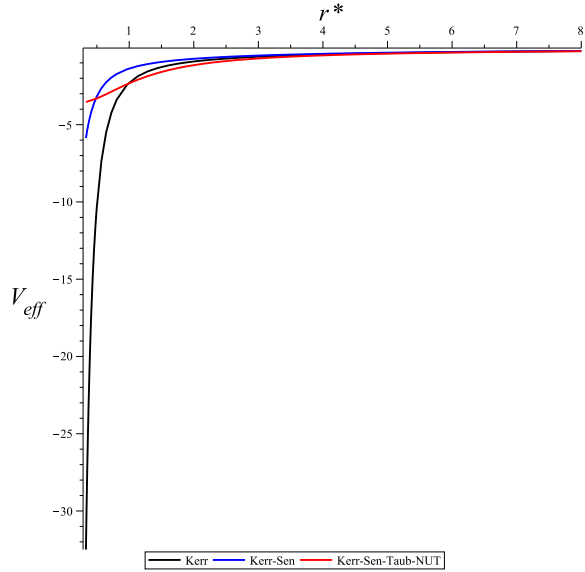}
	\end{center}
	\caption{Effective potentials for timelike particle with $E=1$ and $L^* = 1$ around a collapsing mass with rotational parameter $a^* = 0.5$. The incorporated electric charge parameter $b^* = 0.2$, and NUT parameter $l^*=1$.}\label{Fig.KKSKSTN}
\end{figure}

Also, quite frequently the complexity of full expression for effective potential $V_{r,{\rm{eff}}}$ does not allow us to easily extract the desired informations from the system. Hence, we could rely on the numerical plots for $V_{r,{\rm{eff}}}$ to help in giving insights for some particular questions, despite the answer cannot be general. For example, by looking at the $V_{r,{\rm{eff}}}$ one can tell whether a test particle can travel from infinity to the vicinity of black holes. In this section, we do not aim to extract all the information we can get from the $V_{r,{\rm{eff}}}$ plots, but rather to verify the non-singular value of Kretschmann scalar in KSTN spacetime at $r=0$. We also like to learn the effect of non-vanishing NUT parameter to the profile of $V_{r,{\rm{eff}}}$. Therefore, limiting ourself to the case of timelike test particle only would be adequate, and the corresponding effective potential is (\ref{Vgenr}) with $\delta = -1$. In fig. \ref{Fig.KKSKSTN}, we can observe that  the behavior of effective potential in the presence of NUT parameter $l$ is distinguishable from the vanishing $l$ counterparts, i.e. Kerr and Kerr-Sen. The effective potentials for Kerr and Kerr-Sen are singular at physical singularity $r=0$, but it is not for KSTN case. This resembles the analogous property of Kerr-Newman-Taub-NUT spacetime, which also differs from the case where Taub-NUT parameter is absent, i.e. the family of Kerr-Newman black holes. The finiteness of effective potential for the case represented by fig. \ref{Fig.KKSKSTN} of KSTN spacetime can be understood from the non-singular Kretschmann scalar (\ref{RR}) evaluated at $r=0$. Nevertheless, the plots $V_{r,{\rm{eff}}}$ for each cases in fig. \ref{Fig.KKSKSTN} start to coincide at the larger radius. 

\begin{figure}
	\begin{center}
		\includegraphics[scale=0.4]{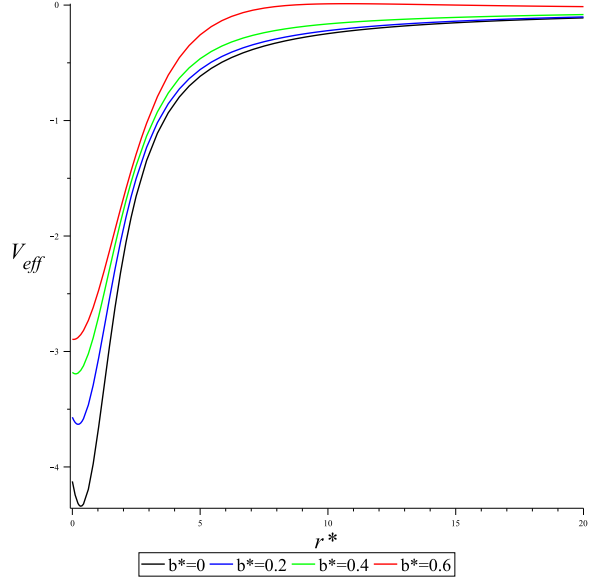}
	\end{center}
	\caption{Effective potentials for timelike particle with $E=1$, $L^* = 2$, $l^*=2$, and the rotational parameter $a^* =0.5$. The electric field $b^*$ varies from $0$ to $0.6$.} \label{Fig.VeffBvar}
\end{figure}

Now let us see what happens when the electric charge parameter $b$ increases, as shown in fig. \ref{Fig.VeffBvar}. Here we observe that near $r=0$ the plots is steeper as $b^*$ decreases\footnote{For economical reason we use the starred notation for all quantities in consideration to denote their ratio to black hole mass, namely $b^*\equiv b/M$, $L^*\equiv L/M$, $r^*\equiv r/M$, $l^*\equiv l/M$, and $a^*\equiv a/M$ as in fig. \ref{fig.ISCO}. }. Alternatively, one can state that the gravitational attraction is stronger when $b$ vanishes. Note that our test timelike particle is electrically neutral, hence Coulomb interaction does not exist between the object that curves the spacetime and the test particle. This finding is similar to that of Kerr-Sen case, where the effective potential plots for a neutral test particle raise as one increases the electric charge of the black hole. 

\begin{figure}
	\begin{center}
		\includegraphics[scale=0.4]{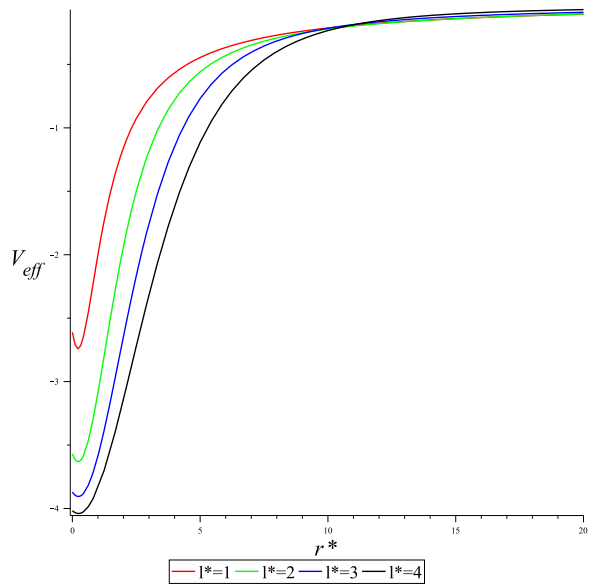}
	\end{center}
	\caption{Effective potentials for timelike particle with $E=1$, $L^* = 2$, $b^*=0.2$, and the rotational parameter $a^* =0.5$. The NUT parameter $l^*$ varies from $1$ to $4$.} \label{Fig.Vefflvar}
\end{figure}

\begin{figure}
	\begin{center}
		\includegraphics[scale=0.4]{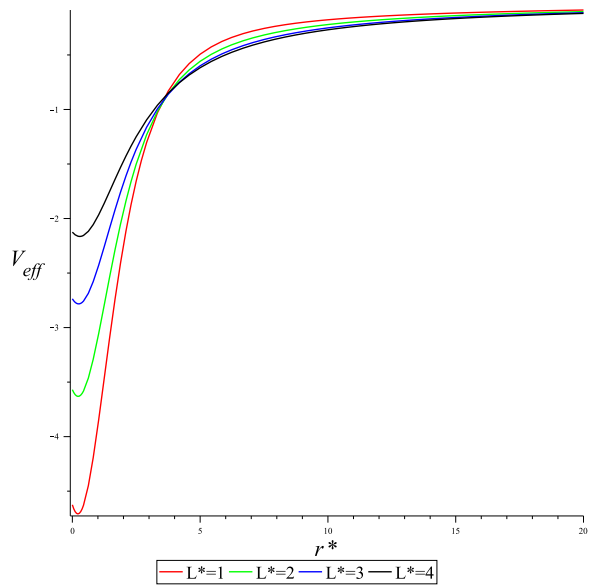}
	\end{center}
	\caption{Effective potentials for timelike particle with $E=1$, $l^* = 2$, $b^*=0.2$, and the rotational parameter $a^* =0.5$. The angular momentum $L^*$ varies from $1$ to $4$.} \label{Fig.VeffLLvar}
\end{figure}

On the other hand, as shown in fig. \ref{Fig.Vefflvar}, we find the plots is deeper for larger values of Taub-NUT parameter $l$ near the center of radius. In this region, the gravitational attraction is stronger as $l$ increases. Similar situations exist in the case of Einstein-Maxwell theory, for example in the plots presented in \cite{Pradhan:2014zia}, but the lines start to coincide at larger radius. Another similarity between KSTN and KNTN cases for timelike particle geodesics is the plots of $V_{r,{\rm{eff}}}$ for different angular momentum $L$. Near the center of coordinate $r=0$, $V_{r,{\rm{eff}}}$ is steeper for lower angular momentum, but the curves start to intersect at some distance. Moreover, from fig. \ref{Fig.VeffLLvar} we can learn that the $V_{r,{\rm{eff}}}$ curves for different angular momentum intersect at a point which is the positive root of quadratic $r^*$ equation\footnote{One can also find an equation dictating the intersection points between curves in fig. \ref{Fig.Vefflvar}, which turn out to be some quintic polynomials in $r^*$. Hence, we do not think it is necessary to express the equation here.}
\[
\left( {4El^*  + \left( {b^*  - 1} \right)\left( {p + q} \right)} \right)\left( {r^* } \right)^2  + 2\left( {b^* - 1} \right)\left( {\left( {b^* - 1} \right)\left( {p + q} \right) + 2E\left( {2l^*  + a^* } \right)} \right)r^* 
\]
\be 
- \left( {l^* } \right)^2 \left( {\left( {b^* - 1} \right)\left( {p + q} \right) + 4E\left( {l^*  + a^* } \right)} \right) = 0\,.
\ee 
In equation above, $p$ and $q$ are the values of $L^*$ under consideration. 

\section{Conclusion}\label{sec.CONC}

In this paper, we obtain a solution describing rotating charged mass with NUT parameter in the low energy limit of heterotic string theory, namely the Kerr-Sen-Taub-NUT spacetime. This solution resembles the Kerr-Newman-Taub-NUT spacetime in Einstein-Maxwell theory. As we expect, there exist several similarities between the two solutions, for example the non-singular squared Kretschmann scalar at $r=0$ and the profile of $V_{\rm eff}$ plots as NUT parameter $l$ increases. 

In section \ref{sec.masscharge}, we also study the conserved charges associated to the new solution. The mass and angular momentum are calculated using Barnich-Brandt method, where the obtained mass is in agreement to that computed using standard Komar integral. However, we find a contribution from the NUT parameter $l$ in the angular momentum, which resembles the cases of Kerr-Taub-NUT or Kerr-Newman-Taub-NUT spacetimes. The electric charge is the same to that of Kerr-Sen black holes. 

Related to the equatorial circular geodesics in both timelike and null considerations given in section \ref{sec.CirGEO}, we learn that such motions cannot exist in Kerr-Sen-Taub-NUT. This conclusion is obvious in the null geodesic case, but is taken based on several numerical evaluations in the timelike consideration. However, we are convinced that any numerical results similar to the plots in fig. \ref{fig.ISCO} lead to the same conclusion for any possible numerical values of $b$. Our results presented in section \ref{sec.CirGEO} agree to the prediction made in \cite{Jefremov:2016dpi}, namely circular geodesics in spacetime with NUT parameter do not exist in equatorial plane.

There are future potential projects that can be done related to Kerr-Sen-Taub-NUT solution presented in this paper. First is the non-equatorial circular motions in Kerr-Sen-Taub-NUT spacetime. This is interesting since it may have some astronomical interests related to the observation of black hole image \cite{Akiyama:2019cqa}. Since in this work we point out that circular null geodesics cannot take place on equatorial plane in Kerr-Sen-Taub-NUT spacetime, one might wonder where they can exist.
Also, here we have not obtained the separable Hamilton-Jacobi equation for test particles in Kerr-Sen-Taub-NUT spacetime. The constraints that yields $\dot r=0$ and $\dot x = 0$ for the test particle geodesics can be derived from this equation, and in general are independent each other. We aim to include this project in our future work. Another interesting problem that can be worked out is the Banados-Silk-West effect \cite{Banados:2009pr,Zakria:2015eua} in Kerr-Sen-Taub-NUT spacetime. Indeed, the latter project is related to the previous one, namely obtaining the separable Hamilton-Jacobi equation for test particles in the spacetime. 

\section*{Acknowledgement}
 
This research is supported by LPPM-UNPAR under the grant no. PL72018027. I thank my colleagues from physics dept. UNPAR for their encouragements.

\end{document}